\documentclass[prb,twocolumn,showpacs,preprintnumbers,amsmath,amssymb,superscriptaddress]{revtex4}

\usepackage{graphicx}
\usepackage{dcolumn}
\usepackage{bm}

\begin{document}


\title{Pressure-induced change of the pairing symmetry in superconducting CeCu$_2$Si$_2$}

\author{E. Lengyel} \email{lengyel@cpfs.mpg.de} \affiliation{%
Max Planck Institute for Chemical Physics of Solids, 01187 Dresden, Germany
}%
\author{M. Nicklas} \affiliation{%
Max Planck Institute for Chemical Physics of Solids, 01187 Dresden, Germany
}%
\author{H. S. Jeevan}
 \altaffiliation[Present address: ]{I.\ Physik.\ Institut, Georg-August-Universit\"{a}t
G\"{o}ttingen, 37077 G\"{o}ttingen, Germany} \affiliation{%
Max Planck Institute for Chemical Physics of Solids, 01187 Dresden, Germany
}%
\author{G. Sparn}
 \altaffiliation[Present address: ]{Max Planck Institute for Nuclear Physics, 69117 Heidelberg, Germany} \affiliation{%
Max Planck Institute for Chemical Physics of Solids, 01187 Dresden, Germany
}%
\author{C. Geibel} \affiliation{%
Max Planck Institute for Chemical Physics of Solids, 01187 Dresden, Germany
}%
\author{F. Steglich} \affiliation{%
Max Planck Institute for Chemical Physics of Solids, 01187 Dresden, Germany
}%
\author{Y. Yoshioka} \affiliation{%
Division of Materials Physics, Department of Materials Engineering Science,
Graduate School of Engineering Science,
Osaka University, Toyonaka 560-8531, Japan
}%
\author{K. Miyake} \affiliation{%
Division of Materials Physics, Department of Materials Engineering Science,
Graduate School of Engineering Science,
Osaka University, Toyonaka 560-8531, Japan
}%

\date{\today}

\begin{abstract}
Low-temperature $(T)$ heat-capacity measurements under hydrostatic
pressure up to $p \approx 2.1$~GPa have been performed on
single-crystalline CeCu$_2$Si$_2$. A broad superconducting (SC)
region exists in the $T-p$ phase diagram. In the low-pressure region
antiferromagnetic spin fluctuations and in the high-pressure region
valence fluctuations had previously been proposed to mediate Cooper pairing. We
could identify these two distinct SC regions. We found different thermodynamic
properties of the SC phase in both regions, supporting the proposal
that different mechanisms might be implied in the formation of
superconductivity. We suggest that different SC order parameters are characterizing the two distinct SC regions.

\end{abstract}

\maketitle

\begin{figure}[b!]
\includegraphics[angle=0,width=7cm,clip]{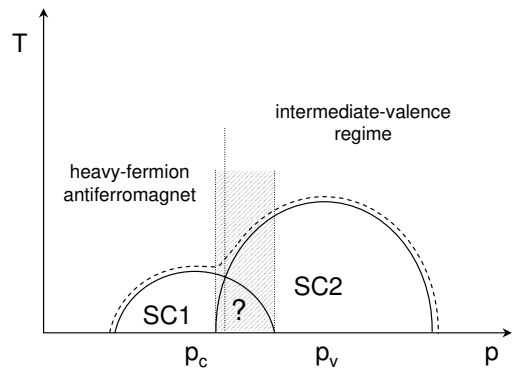}
\caption{\label{fig1} Schematic $T - p$ phase diagram of
CeCu$_2$Si$_2$. The dashed line depicts the shape of the SC phase
line observed experimentally. $p_c$ and $p_v$ indicate the critical
pressures for the magnetic and the valence instability,
respectively.}
\end{figure}

The ongoing interest in unconventional, i.e., non-$s$-wave, superconductors was initiated 30 years ago by the discovery of superconductivity in the heavy-fermion (HF) metal CeCu$_2$Si$_2$.\cite{steg79} While for conventional (BCS) superconductors a very low concentration of magnetic impurities is generally detrimental to superconductivity, for CeCu$_2$Si$_2$ 100at\% of magnetic Ce$^{3+}$ ions turned out to be prerequisite to form the SC phase:\cite{steg79} The non-magnetic reference compound LaCu$_2$Si$_2$ is not a superconductor,\cite{steg79} and doping with a small amount of non-magnetic impurities was found to suppress the SC state completely.\cite{spil83} Because of their small effective Fermi velocity, the heavy quasiparticles forming the Cooper pairs in CeCu$_2$Si$_2$ cannot escape their own ``polarization cloud" which discards the BCS-type electron-phonon coupling mechanism.\cite{steg79} Soon after the discovery of HF superconductivity, magnetic couplings were considered to mediate the pairing in these materials.\cite{schm86} As early as 1986, antiferromagnetic (AF) spin fluctuations, including those at low frequencies near a spin-density-wave (SDW) instability (or quantum critical point, QCP) were proposed to act as SC glue in HF metals.\cite{scal86} The pressure-induced superconductor CePd$_2$Si$_2$ may be considered of prototype for this type of superconductors:\cite{gros96} It exhibits a very narrow ``dome" of superconductivity centered around its QCP at a critical pressure $p_c \approx 2.8$~GPa and, further on, shows pronounced non-Fermi-liquid (NFL) behavior in its low-temperature normal state.\cite{gros96} Remarkably, in CeCu$_2$Si$_2$ superconductivity extends well beyond the AF
instability, suggesting that a mechanism, other than AF spin
fluctuations, might be involved in the formation of the Cooper pairs
in the high-pressure region. There, valence fluctuations were
supposed to mediate the formation of superconductivity.\cite{bell84, jacc99} The extended SC state of CeCu$_2$Si$_2$,
schematically depicted in Fig.~\ref{fig1}, results from the merging
of two distinct SC regions:\cite{yuan03} The one on the
low-pressure side (SC1) appears to be similar to that observed in
other NFL superconductors, like CePd$_2$Si$_2$,\cite{gros96}
while in the high-pressure region,
 a novel type of SC state (SC2) seems to form. Even though valence-fluctuation mediated superconductivity has been predicted theoretically (e.g.
Ref.~\onlinecite{onis00, mont04}), its experimental observation is so far
limited to the CeCu$_2$(Si$_{1-x}$Ge$_x$)$_2$ family.\cite{jacc99,
yuan03} In these materials the two instabilities, i.e., an AF one
at low pressure and a low-lying critical end point of the first-order
valence-transition line at elevated pressure,\cite{yuan06} are sufficiently
separated in order to be distinguishable and, at the same
time, show up in an experimentally accessible pressure range. While superconductivity in the regions SC1 and SC2 has been presumed to be
mediated by AF spin fluctuations and critical valence fluctuations, respectively,\cite{bell84, jacc99, yuan03} the pairing mechanism in the crossover region (hatched area in
Fig.~\ref{fig1}) from the HF to the intermediate valence (IV) state
is still a matter of discussion: While both types of fluctuations may be
involved together in forming the SC state on the one hand, a first-order
transition line might separate the two distinct regions of superconductivity on the other.

In this paper, we study the thermodynamic properties in SC CeCu$_2$Si$_2$ by specific-heat experiments under pressure. For the present study  we have chosen a CeCu$_2$Si$_2$ single crystal of stoichiometric composition, in which superconductivity expels SDW order at low magnetic field, but where the SDW is recovered in an overcritical magnetic field for superconductivity (``$A/S$-type" CeCu$_2$Si$_2$).\cite{steg96} We find
different thermodynamic properties in the two distinct SC regions,
SC1 and SC2. This hints at a pressure-induced change of the pairing
symmetry.

\begin{figure}[t!]
\includegraphics[angle=0,width=8cm,clip]{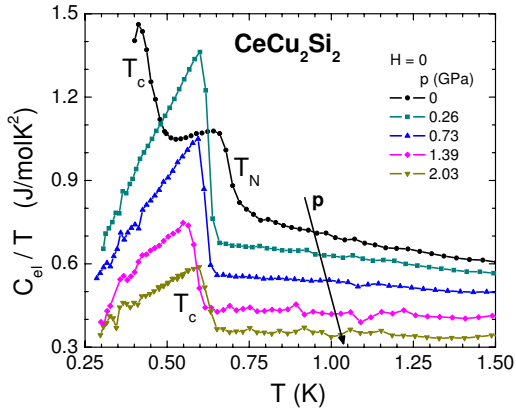}
\caption{\label{fig2} (Color online) Low-temperature $C_{\rm el}(T)/T$ versus $T$
of $A/S$-type CeCu$_2$Si$_2$ at $H = 0$ for pressures as indicated
in the figure.}
\end{figure}

The $A/S$-type CeCu$_2$Si$_2$ single crystal was grown
in an aluminum-oxide crucible by a modified Bridgman technique, using
Cu excess as flux medium. Powder X-ray diffraction patterns
confirmed the proper tetragonal ThCr$_2$Si$_2$ structure with
lattice parameters $a = 0.4099$~nm and $c = 0.9923$~nm at 295~K.
Heat-capacity measurements under hydrostatic pressure have been
performed in a single-shot $^3$He evaporation cryostat by employing
a compensated quasi-adiabatic heat-pulse technique. In addition, the
SC transition in CeCu$_2$Si$_2$ was detected through magnetocaloric
and a.c.-susceptibility measurements on the same sample and at the
same pressures. A single piece of $A/S$-type CeCu$_2$Si$_2$ weighing
about $m \approx 0.4$~g was used for the experiments. Its residual
resistivity was $\rho_0 \approx 10~\mu\Omega{\rm cm}$, indicating a
good sample quality.\cite{rauc87} The magnetic field was always
applied parallel to the $c$-axis. The measurements at low pressures
($p < 1.1$~GPa) were carried out in a CuBe piston-cylinder pressure
cell, while for the high-pressure range ($p \geq 1.1$~GPa) a double
layer NiCrAl-CuBe type piston-cylinder pressure cell was utilized.
For the entire experiment, Flourinert FC72 was used as pressure
transmitting medium. A piece of tin served as pressure gauge. For
the whole pressure range, the electronic specific heat ($C_{\rm
el}$) was obtained by subtracting the ambient pressure lattice
specific heat of the isostructural non-magnetic reference compound
LaCu$_2$Si$_2$.\cite{hellup}

The temperature dependence of the low-temperature $C_{\rm
el}$ of CeCu$_2$Si$_2$ for selected pressures is shown in
Fig.~\ref{fig2}. Characteristic for $A/S$-type CeCu$_2$Si$_2$, two consecutive phase
transitions can be observed at $p = 0$, an upper one at $T_{\rm N} \approx
0.69$~K to an incommensurate SDW order
[Ref.~\onlinecite{stoc04}] and a lower one marking the onset of superconductivity at $T_{\rm c} \approx
0.46$~K. Upon increasing pressure, the AF order is gradually
suppressed while superconductivity is stabilized. Above $p = 0.07$~GPa no
anomaly indicating the onset of AF order can be observed anymore.
In the normal state, $C_{\rm el}$ of CeCu$_2$Si$_2$ decreases with increasing
pressure in the entire pressure range. This is expected for Ce-based HF systems, where application of pressure
leads to an increase of the hybridization strength between the
Ce-$4f$ and the conduction electrons and hence to a decrease
of the effective mass of the quasiparticles.

\begin{figure}
\includegraphics[angle=0,width=8cm,clip]{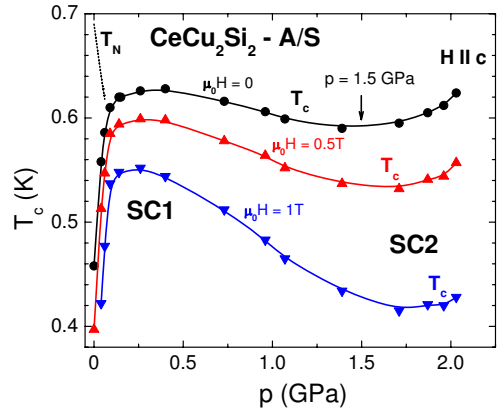}
\caption{\label{fig3} (Color online) Pressure dependence of $T_{\rm c}$ for
different values of $\mu_0 H$. The pressure dependence of $T_{\rm
N}$ at $H = 0$ is shown by the dotted line. The arrow
indicates the estimated border between the
low-pressure (SC1) and the high-pressure (SC2) SC regions at $H = 0$.}
\end{figure}

Figure \ref{fig3} displays the evolution of $T_{\rm c}$ as function
of pressure for different magnetic fields as obtained from the
heat-capacity measurements. With increasing pressure, a steep initial rise of $T_{\rm c}(p)$ is followed by a pronounced maximum and, at elevated pressure, a shallow minimum. For $H = 0$, $T_{\rm c,max} \approx 0.63$~K at $p \approx 0.4$~GPa and $T_{\rm c,min} \approx 0.59$~K at $p \approx 1.5$~GPa. We use the pressure value of $T_{\rm c,min}$ to delineate the border between the two regions SC1 and SC2. With
increasing magnetic field the $T_{\rm c}$ values become reduced and
$T_{\rm c,min}$ shifts to higher pressures, suggesting
an increasing separation between the two SC regions. As can be seen in
Fig.~\ref{fig3}, superconductivity in the SC2 region is suppressed
more efficiently by the magnetic field than in the SC1 region. At
$\mu_0 H = 2$~T, superconductivity still exists in a very narrow
pressure range at low pressures, while in the high-pressure region
there is no superconductivity up to $p \approx 2.1$~GPa. Here, the
upper-critical field, $\mu_0 H_{\rm c2}(0)$, is smaller than 1.5~T.
In the low-temperature normal state ($T < 1$~K, $\mu_0 H = 2$~T),
$C_{\rm el}(T)/T = {\rm const.}~(\approx 0.4$~J/(molK$^2$)) at $p >
1.5$~GPa, indicating a moderately heavy Landau Fermi-liquid state.
This leads us to conclude that in this pressure and magnetic
field range $A/S$-type CeCu$_2$Si$_2$ is situated far away from a
QCP.

\begin{figure}[t!]
\includegraphics[angle=0,width=7.5cm,clip]{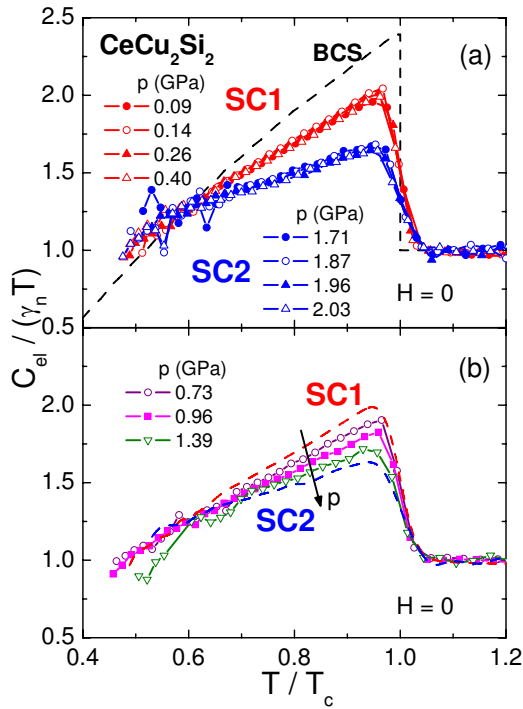}
\caption{\label{fig4} (Color) Normalized electronic specific-heat data
obtained on CeCu$_2$Si$_2$ under pressure. The dashed line in panel
$a$ corresponds to the theoretically calculated dependence for the
case of a conventional BCS-type superconductor. The two dashed lines
in panel $b$ serve as references and they reproduce the data for the
two distinct groups presented in panel $a$.}
\end{figure}

\begin{figure}[t!]
\includegraphics[angle=0,width=7.5cm,clip]{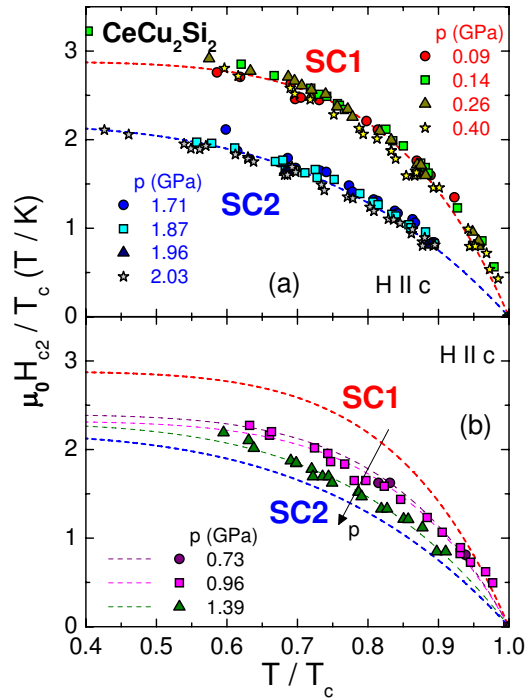}
\caption{\label{fig5} (Color) $\mu_0 H_{\rm c2}(T)/T_{\rm c}$ versus
$T/T_{\rm c}$ for different pressures obtained for $H \parallel c$.
Lines are guides to the eyes. The red and blue line from panel $b$
reproduce the corresponding lines from panel $a$ and serve as
references for the regions SC1 and SC2 presented in panel $a$.}
\end{figure}

Figure \ref{fig4} depicts the evolution of the $H = 0$ normalized
low-temperature electronic specific heat under pressure. The data
are presented as $C_{\rm el}(T)/(\gamma_{\rm n}T)$ versus $T/T_{\rm
c}$, where $\gamma_{\rm n}$ is $C_{\rm el}/T \mid_{T = T^+_{\rm c}}$
in the normal state. The specific-heat data obtained on SC CeCu$_2$Si$_2$ does
not follow the BCS prediction (dashed line), and the $\Delta
C_{\rm el}/(\gamma_{\rm n}T_{\rm c})\mid_{T = T_{\rm c}}$ ratio
exhibits values smaller than the BCS value of 1.43. A quasi-linear
temperature dependence of $C_{\rm el}(T)/T$ can be observed at
$0.5T_{\rm c} < T < T_{\rm c}$ for all pressures above $0.09$~GPa. A comparison with numerical
calculations of $C_{\rm el}(T)/T$ [Ref.~\onlinecite{hass93}] suggests that the SC
state in CeCu$_2$Si$_2$ has an unconventional nature and is
characterized by a gap function having line nodes.

The normalized specific-heat data, $C_{\rm el}(T)/(\gamma_{\rm n}T)$
versus $T/T_{\rm c}$,  fall on a single curve for pressures
$0.09~{\rm GPa} \leq p \leq 0.4$~GPa; in the same way, the data in
the high-pressure range, $1.71~{\rm GPa} \leq p \leq 2.03$~GPa,
collapse also on a single (but different) curve as can be seen in
Fig.~\ref{fig4}a. Fig.~\ref{fig4}b displaying the data for the intermediate pressure
range $0.73~{\rm GPa} \leq p \leq 1.39$~GPa reveals a gradual shift
of the data from the low-pressure to the high-pressure
scaling curve. Fig.~\ref{fig5} presents the normalized
upper-critical field $\mu_0 H_{\rm c2}(T)/T_{\rm c}$ as function of
the normalized temperature $T/T_{\rm c}$. These data display a
similar pressure evolution as observed in the case of $C_{\rm el}(T)/(\gamma_{\rm n}T)$: Two distinct scaling curves are found for
the two SC regions, and the data in the intermediate pressure range
shift gradually on increasing pressure from the scaling curve
corresponding to region SC1 to the one corresponding to region SC2.
These findings highlight that the SC order parameters in regions SC1 and SC2 are different. The
continuous evolution of the data from SC1 to SC2 favors an overlap
region between SC1 and SC2 where a smooth crossover takes place,
rather than a first-order transition line between SC1 and SC2.

At $p \geq 0.09$~GPa, the values estimated for the Pauli-limiting
field are slightly smaller than those experimentally obtained for
the upper-critical field, while for the orbital-limiting field we
estimate values 3 to 4 times larger than $H_{\rm c2}(0)$. This
proves that $H_{\rm c2}(T)$ is strongly Pauli limited in
the pressure range $0.09~{\rm GPa} \leq p < 2.1$~GPa, indicating
a SC order parameter of even parity, consistent with $d$-wave pairing symmetry as we will discuss in the following.

Our conclusion of different SC order parameters in CeCu$_2$Si$_2$ at low and high pressures is corroborated by theoretical considerations. The experimental results presented in Figs.~\ref{fig4}a and
\ref{fig5}a show that $\Delta C_{\rm el}/(\gamma_{\rm n}T_{\rm
c})\mid^{\rm SC1}_{T = T_{\rm c}}/\Delta C_{\rm el}/(\gamma_{\rm
n}T_{\rm c})\mid^{\rm SC2}_{T = T_{\rm c}} \approx 1.6$ and $({\rm
d}H_{{\rm c}2}/{\rm d}T)^{\rm SC1}_{T=T_{\rm c}}/({\rm d}H_{{\rm
c}2}/{\rm d}T)^{\rm SC2}_{T=T_{\rm c}} \approx 1.9$. A theoretical analysis shows that even
within the manifold of ``\textit{d}"-symmetry the above mentioned ratios can be
different from 1 for different symmetries of pairing:\cite{yoshup} By choosing the $d_{x^{2}-y^{2}}$ and the
$d_{xy}$ type pairings for the SC1 and SC2 region, respectively, one estimates theoretically
$\Delta C_{\rm el}/(\gamma_{\rm n}T_{\rm
c})\mid^{d_{x^{2}-y^{2}}}_{T = T_{\rm c}}/\Delta C_{\rm el}/(\gamma_{\rm
n}T_{\rm c})\mid^{d_{xy}}_{T = T_{\rm c}} \approx 1.6$ and $({\rm
d}H_{{\rm c}2}/{\rm d}T)^{d_{x^{2}-y^{2}}}_{T=T_{\rm c}}/({\rm d}H_{{\rm
c}2}/{\rm d}T)^{d_{xy}}_{T=T_{\rm c}} \approx 1.8$.\cite{yoshup} The good agreement between the experimental results and the theoretical
estimation for these quantities suggests that $d_{x^2-y^2}$ pairing is realized at the lower pressure side (region SC1) and $d_{xy}$ pairing is realized at
the higher pressure side (region SC2). It is reasonable that
$d_{x^2-y^2}$ pairing is realized in the lower pressure region
where the AF fluctuations develop due to AF quantum criticality
around ambient pressure, because the AF correlations among $f$
electrons at adjacent sites in the basal plane are expected to
promote the pairing with $d_{x^2-y^2}$ symmetry.\cite{miya86} The
origin of the $d_{xy}$ pairing in the higher pressure region
should be assigned to a SC glue different from AF fluctuations.
As mentioned earlier, a promising candidate may be valence
fluctuations which are enhanced in the higher pressure region around
$p\simeq5$~GPa.\cite{bell84, holm04}

In conclusion, we found different thermodynamic properties in the
two distinct SC regions of CeCu$_2$Si$_2$. Our results support the
previously made suggestion [Ref.~\onlinecite{bell84, jacc99, yuan03, onis00, mont04}] that two
different mechanisms are involved in the formation of the Cooper
pairs in these two regions: In the SC1 state, pairing is likely to be
mediated by AF spin fluctuations; in the high-pressure SC state
(SC2), valence fluctuations are supposed to mediate the formation of
the Cooper pairs. Superconductivity in the low-pressure region is more robust against application of
magnetic field than in the SC2 region as indicated by the larger
upper-critical fields. Further on, we observed distinct scaling laws of
$C_{\rm el}(T)/(\gamma_{\rm n}T)$ versus $T/T_{\rm c}$ and of $\mu_0
H_{\rm c2}(T)/T_{\rm c}$ versus $T/T_{\rm c}$ in the two different
SC regions. Therefore, we suggest the existence of different SC
order parameters in SC1 and SC2. A theoretical
analysis of our data proposes $d_{x^2-y^2}$ type
Cooper-pairing for the SC1 region and  $d_{xy}$ type
pairing for the SC2 region. The existence of different SC order
parameters is highly consistent with the different mechanisms supposed to be
implied in the formation of Cooper pairs in CeCu$_2$Si$_2$. We
find a smooth crossover from the SC1 to the SC2 region. Thus, this
crossover region should be characterized by a SC state where both AF
spin and valence fluctuations are involved together in the Cooper
pairing. However, for a precise experimental determination of the SC order
parameters in the low- and high-pressure regimes, field-angle dependent specific-heat experiments at low temperatures have to be performed in the future.

\begin{acknowledgments}
This work was partly supported by the DFG under the auspices of both the SFB 463
and the Research Unit 960. The work at Osaka University was supported by a Grant-in-Aid for
Scientific Research (19340099) from the Japan Society for the
Promotion of Science. One of us (Y. Y.) is supported by the G-COE
program (G10) from the Japan Society for the Promotion of Science.
\end{acknowledgments}



\bibliography{bibccs1}

\begin{thebibliography}{19}
\bibitem[{\citenamefont{Yoshioka and Miyake}(in preparation)}]{yoshup}
\bibinfo{author}{\bibfnamefont{Y.}~\bibnamefont{Yoshioka}} \bibnamefont{and}
  \bibinfo{author}{\bibfnamefont{K.}~\bibnamefont{Miyake}} (\bibinfo{year}{in
  preparation}), \bibinfo{note}{the analysis is based on a Ginzburg-Landau
  approach and takes into account the specific Fermi surface of
  CeCu$_2$Si$_2$.}
\end{thebibliography}

\end{document}